\documentclass[conference]{IEEEtran}
\IEEEoverridecommandlockouts
\usepackage{cite}
\usepackage{amsmath,amssymb,amsfonts}
\usepackage{algorithmic}
\usepackage{graphicx}
\usepackage{textcomp}
\usepackage{xcolor}
\def\BibTeX{{\rm B\kern-.05em{\sc i\kern-.025em b}\kern-.08em
    T\kern-.1667em\lower.7ex\hbox{E}\kern-.125emX}}
\begin{document}

\title{Performance Evaluation of Hashing Algorithms on Commodity Hardware\\

}

\author{\IEEEauthorblockN{Marut Pandya }
\IEEEauthorblockA{\textit{Data Intensive Computing, CS554} \\
\textit{Department of Computer Science, Illinois Institute of Technology}\\
Chicago, Illinois, USA \\
mpandya6@hawk.iit.edu}

}

\maketitle

\begin{abstract}
Hashing functions, which are created to provide brief and erratic digests for the message entered, are the primary cryptographic primitives used in blockchain networks. Hashing is employed in blockchain networks to create linked block lists, which offer safe and secure distributed repository storage for critical information. Due to the unique nature of the hash search problem in blockchain networks, the most parallelization of calculations is possible. This technical report presents a performance evaluation of three popular hashing algorithms: Blake3, SHA-256, and SHA-512. These hashing algorithms are widely used in various applications, such as digital signatures, message authentication, and password storage. It then discusses the performance metrics used to evaluate the algorithms, such as hash rate/throughput and memory usage.
The evaluation is conducted on a range of hardware platforms, including desktop and VMs. The evaluation includes synthetic benchmarks. The results of the evaluation show that Blake3 generally outperforms both SHA-256 and SHA-512 in terms of throughput and latency. However, the performance advantage of Blake3 varies depending on the specific hardware platform and the size of the input data.
The report concludes with recommendations for selecting the most suitable hashing algorithm for a given application, based on its performance requirements and security needs. The evaluation results can also inform future research and development efforts to improve the performance and security of hashing algorithms.

\end{abstract}


\section{Introduction}
Blockchain technology is a huge technological advancement with enormous promise, and both the government and business sectors are paying close attention to its uses. Although Bitcoin was the first blockchain application, its use has since expanded to other fields such as biomedicine, supply chain management, and smart contract registration. Blockchain operates as a distributed and decentralized database that consists of a sequence of blocks containing a list of accumulated transactions. Each block comprises three primary sections: data, hash block, and previous hash block. The hash block acts as a unique identifier for each block and is obtained by using mathematical rules to encrypt information contained in the block. Moreover, each block contains the hash of the preceding block, creating a chain of blocks. Any unauthorized modifications to the information of a block will alter its hash number, making the block invalid for the subsequent blocks. A systematic survey on the application of blockchain technology in various fields has been conducted. Blockchain technology has enormous potential to store ten percent of global GDP by 2027, according to the World Economic Forum survey. The first introduction of blockchain technology was by a group of researchers, Haber and Stornetta in 1990, but it was not until the establishment of Bitcoin by Satoshi Nakamoto in 2008 that it gained widespread applications.

In this technical report, Ievaluate the performance of three popular hashing algorithms - BLAKE3, SHA-256, and SHA-512 - in the context of blockchain applications. Iexamine it's performance. The contribution of this work is an experimental evaluation of hashing algorithms on two hardware, Macbook pro and t2.2xLarge ec2 instance. The results demonstrate that there are platform-specific differences to hash
function performance. The report also includes the implementation details of the benchmarks, machine configuration of the two hardwares and results. 

\section{Background}
Blockchains are distributed ledgers that store data in blocks that are linked together using cryptographic hashes. Hashing algorithms are used to generate a unique identifier for each block, which is included in the next block to form a chain. This ensures that any modification to a block will result in a different hash value, thus preserving the integrity of the data stored on the blockchain.

BLAKE3, SHA-256, and SHA-512 are popular hashing algorithms used in various blockchain implementations. SHA-256 is widely used in Bitcoin, while Ethereum uses a modified version of SHA-3 called Keccak-256. BLAKE3 is a relatively new algorithm that has gained attention for its speed and security properties.

A very recent hashing method, BLAKE3, was unveiled in 2020. It can generate hash values of different widths, ranging from 256 to 512 bits, and is intended to be faster and more secure than its forerunners. In order to provide parallelism and optimal memory utilization, BLAKE3 adopts a tree-based methodology, making it a suitable contender for a variety of applications.
As opposed to this, SHA-256 and SHA-512 are hashing algorithms that are a part of the SHA-2 family and are frequently employed in security-related applications. When compared to SHA-512, SHA-256 generates hash values that are 256 bits instead of 512 bits. Both algorithms are built using the Merkle-Damgard construction and rely on a compression function that works with fixed-size message blocks.

\section{Motivation}
The motivation for evaluating the performance of BLAKE3, SHA-256, and SHA-512 in the context of blockchain is driven by the importance of hashing algorithms in blockchain technology, the emergence of new hashing algorithms, the increasing complexity of blockchain applications, and the impact on blockchain mining.  The importance of hashing algorithm, Increasing the complexity of the blockchain and impact on bitcoin mining  are few of the motivation behind doing these evaluation as well because hashing algorithms play crtical role in blockchain technology, and as this technology is evolving, application that rely on this are getting more complex and hash rate is a crucial factor in determine the success of a mining operation. By evaluating these algorithms, Igive the valuable insights that can help blockchain developers and researchers make informed decisions. 

The motivation for this study also comes from the recent development of the BLAKE3 hashing algorithm, which has been claimed to be faster and more secure than its predecessors. The creators of BLAKE3 have presented it as a high-performance hashing algorithm[Fig. 1] that can operate efficiently on modern CPUs and GPUs, making it well-suited for use in a variety of applications, including blockchain technology. Given the rapid evolution of the field of blockchain and the increasing complexity of blockchain applications, it is important to evaluate the performance of BLAKE3 compared to other established hashing algorithms like SHA-256 and SHA-512 to determine if it can improve the efficiency and security of blockchain networks

\begin{figure}[htbp]
\centerline{\includegraphics{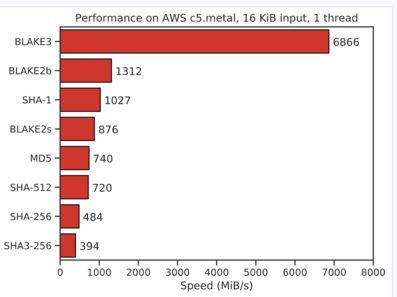}}
\caption{performance graph from Blake3 paper}
\label{fig}
\end{figure}



\section{Proposed Solution}
To study the hashing algorithms better, studied research papers, white papers and design experiments to measure the performance of the three hashing algorithms.

\subsection{Nature of  the Project}\label{AA}
To evaluate the performance of BLAKE3, SHA-256, and SHA-512, Iconducted a series of empirical studies using a variety of test scenarios. Our experiments were designed to measure the memory usage and hash computation per second for each algorithm on different input sizes and to compare their performance under different processing environments, with multiple threads.  To generate the test data I used the srand()
in function to seed the input message to the hash function. I  then measured the hash rate and execution time of each of these algorithms. my goal was to provide a comprehensive performance evaluation of each algorithm that would enable us to determine which algorithm was best suited.

\subsection{Hardware and Software}

I used a MacBook Pro with a 2 GHz quad core Intel Core i5 processor and 16 GB of RAM as our primary testing environment. Additionally, I utilized an EC2 instance (t2.2x large) with 8 virtual CPUs, 32 GB of memory, and high network and I/O performance to simulate a more powerful processing environment. 
Linux EpycBox with 424 GB memory,  AMD EPYC 7501 32-Core Processor, 128 CPUs.


\subsection{Implementation}

\begin{itemize}
\item \textit{Hash Functions} - They are used to ensure the integrity and security of data stored on the blockchain. A hash function is a mathematical algorithm that takes an input (or message) of any size and produces a fixed-size output known as a hash or message digest. They  are also used in the context of blockchain consensus mechanisms, such as proof of work (PoW) and proof of stake (PoS), to ensure that blocks are validated and added to the blockchain in a secure and trustless manner. They play a critical role in blockchain technology, providing a tamper-evident and immutable ledger of transactions that can be trusted without the need for a central authority.

\end{itemize}

Since I used C programming language to develop my benchmarks, To compile my  benchmarking code, I used the GNU Compiler Collection (GCC), which is a widely-used compiler suite that supports multiple programming languages. Specifically, we used GCC version 13.0.0 to compile our benchmarking codeon Macbook machine(Clang - default compiler).  Additionally, I installed the OpenSSL library, which is a popular cryptographic library that provides a range of cryptographic functions, including hashing. I used OpenSSL version 3.3.6 and installed the associated libssl-dev package to access the necessary functions in our benchmarking code. By using these widely-used and established packages/libraries, I was able to ensure that our benchmarking code was compatible with a range of systems and could generate reliable and accurate performance measurements for BLAKE3, SHA-256, and SHA-512.

I utilized the BLAKE3 reference implementation, which is a publicly accessible open-source library, to implement the BLAKE3 algorithm. To access the required functions for my tests, I used the blake.h header file that is available in the BLAKE3 repository. In order to produce hash values for comparison for SHA-256 and SHA-512, I used the OpenSSL library, which offers a well-known and widely-accepted method. For my tests, we utilized OpenSSL 3.3.6.
I implemented multi threaded benchmarks, and based on the requirement I was able to change the number of threads, iterations and input size. I implemented benchmarks for 1,2,4, and 8 threads. Below figure shows the code snippet: 

\begin{figure}[htbp]
\centerline{\includegraphics{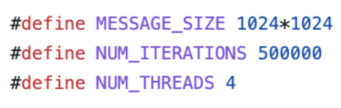}}
\caption{Defines the input size, no. of iteration and total number of threads. (example)}
\label{fig}
\end{figure}

\textit{\textbf{SHA512} - } This benchmark measures the hash rate of SHA-512 hashing algorithm by generating a random input message and repeatedly applying the SHA-512 hash function to it. The code uses multithreading to improve the performance of the computation. The program has several constants defined at the beginning of the file, such as MESSAGE-SIZE, NUM\_ITERATIONS, and NUM\_THREADS.

In the main function, the program first allocates memory ( malloc() ) for the message buffer and an array to store the hashes per second calculated by each thread. Then, it generates a random input message of size MESSAGE\_SIZE. The program then creates NUM\_THREADS threads and assigns a portion of the message buffer to each thread. Each thread executes the worker function, which takes a pointer to a struct containing the message buffer, the number of iterations to perform, the thread ID, and a pointer to the array storing the hashes per second calculated by each thread.

The worker function uses the OpenSSL library's SHA-512 function to hash the input message a number of times specified by the num\_iterations argument. It records the time taken to complete the computation and calculates the hashes per second rate. It then stores this rate in the array pointed to by the hashes\_per\_second argument.

After all threads complete execution, the main function calculates the total hashes per second by summing the values in the hashes\_per\_second array. It also calculates the total time taken by the computation and prints the results.

\begin{figure}[htbp]
\centerline{\includegraphics{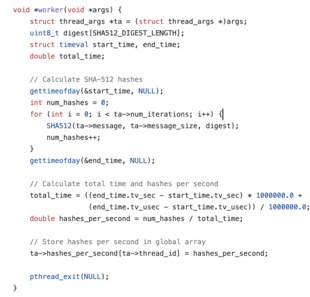}}
\caption{Snippet shows the SHA512 generic worker function}
\label{fig}
\end{figure}

\textbf{\textit{SHA256 - }} This benchmark implements a multithreaded program that measures the hash rate of the SHA-256 algorithm. The program generates a random message of size 1 MB and uses it to perform a specified number of iterations of the SHA-256 hash function. The hash rate is calculated by dividing the number of iterations by the total time taken to complete them.

The program creates a specified number of threads and assigns each thread a portion of the iterations to perform. The number of threads and iterations are defined by the constants NUM\_THREADS and NUM\_ITERATIONS respectively. The message and hash rate variables are shared among all threads through a struct called thread\_data. Each thread takes a pointer to this struct as its argument.

The thread\_func function is the main function executed by each thread. It takes a pointer to the thread\_data struct as its argument, and performs the assigned number of iterations of the SHA-256 hash function on the message provided in the struct. The SHA256 function from the OpenSSL library is used to perform the hash function. The hash rate is calculated by measuring the time taken to perform the iterations and dividing the number of iterations by this time.

The main function generates the random message, creates the threads and assigns each thread its portion of the iterations, and then waits for all threads to complete using pthread\_join. The total time taken to perform all iterations is calculated and the final hash rate is printed.

\begin{figure}[htbp]
\centerline{\includegraphics{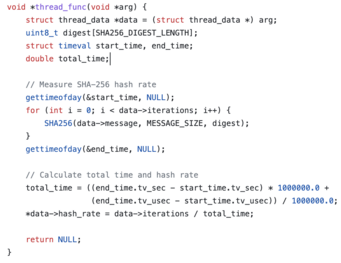}}
\caption{Snippet shows the SHA256 generic hash function}
\label{fig}
\end{figure}

\textbf{\textit{BLAKE3 - }} This benchmark measures the hash rate of BLAKE3, a cryptographic hash function, using multiple threads. The program generates a random input message of size MESSAGE\_SIZE (defined as 1024*1024 bytes) and then divides the number of iterations (NUM\_ITERATIONS) evenly among the number of threads (NUM\_THREADS), creating a struct with the message and number of iterations for each thread.

Each thread runs the hash\_thread function, which initializes a BLAKE3 hasher, updates the hasher with the input message NUM\_ITERATIONS times, finalizes the hasher, and stores the resulting digest in an array. Once all threads have completed their computations, the main thread calculates the hash rate by dividing the number of iterations by the total time it took for all threads to finish.

The program uses the pthread library for multi-threading and the sys/time.h library for measuring time. The hash\_thread function is passed a void pointer to the struct containing the input message and number of iterations as arguments. Once it has completed its computations, it exits using pthread\_exit. The main thread creates NUM\_THREADS threads using pthread\_create, passing a pointer to the struct for each thread. It then waits for all threads to finish using pthread\_join. Once all threads have completed, the main thread calculates the total time and hash rate and prints the results to the console. Finally, the program frees the dynamically allocated memory used for the input message and returns 0.
\begin{figure}[htbp]
\centerline{\includegraphics{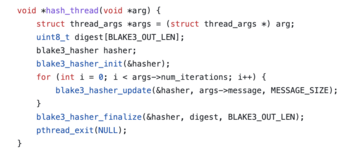}}
\caption{Snippet shows the blake3 generic hash function}
\label{fig}
\end{figure}

\textit{\textbf{Memory Usage} - } Implemented three benchmarks for measuring  memory usage for sha256, blake3, and sha512 respectively.

\textit{for Blake3}, -  memory usage is measured using the getrusage() function, which returns resource usage measures for the calling process. Specifically, it measures the maximum resident set size (in kilobytes) of the process.The ru\_maxrss field of the rusage struct represents the maximum resident set size, which is the maximum amount of physical memory used by the process during its lifetime. This includes both the program's own memory usage and any memory used by its child processes. The value returned by getrusage() is the difference between the maximum resident set size at the end of the program and the maximum resident set size at the start of the program. The ru\_maxrss field is measured in kilobytes, so to print the result in kilobytes, the program subtracts the ru\_maxrss value at the start of the program from the ru\_maxrss value at the end of the program, and then prints the result. Note that the ru\_maxrss value is not a precise measure of the amount of memory used by the program, since it includes memory used by child processes and other factors. However, it provides a reasonable estimate of the program's memory usage.

\textit{for sha512,-} Benchmark program measures the memory usage using the getrusage() function from the <sys/resource.h> header. The getrusage() function returns resource usage measures for the current process or its terminated child processes. In this case, it is being used to obtain resource usage statistics for the current process (RUSAGE\_SELF argument). The struct rusage type is used to store the resource usage statistics returned by getrusage(). In the program, two instances of struct rusage are used: usage\_start and usage\_end. The program calls getrusage() twice: once before the threads are created (getrusage(RUSAGE\_SELF, \&usage\_start)), and once after all the threads have completed (getrusage(RUSAGE\_SELF, usage\_end)).
After both calls to getrusage(), the difference in the ru\_maxrss member of the two struct rusage instances is calculated to determine the memory usage of the program. The ru\_maxrss member represents the maximum resident set size used (in kilobytes) by the process during its lifetime.
The calculation for memory usage is (double) (usage\_end.ru\_maxrss - usage\_start.ru\_maxrss) / 1024, which subtracts the ru\_maxrss value at the start of the program from the ru\_maxrss value at the end of the program, and then converts the result from kilobytes to megabytes by dividing by 1024.

\textit{for sha256 - } In this benchmark memory usage is measured using the getrusage() function, which is part of the POSIX API for resource usage. Specifically, this program is using the RUSAGE\_SELF argument to get information about the resources used by the current process (i.e., the running program). The program calls getrusage(RUSAGE\_SELF, \&start\_usage) to get the initial resource usage information for the process. The program then runs the SHA-256 algorithm on the input message for NUM\_ITERATIONS iterations, using NUM\_THREADS threads. After all threads have completed, the program calls getrusage(RUSAGE\_SELF, \&end\_usage) to get the final resource usage information for the process. The program calculates the memory usage by subtracting the initial ru\_maxrss value from the final ru\_maxrss value, which gives the maximum resident set size (RSS) used by the process during execution.

*** \textit{Maximum resident set size (maxrss) is the maximum amount of physical memory (in kilobytes) that a process has used during its execution. This includes both the memory that was directly allocated by the process and the memory that was shared with other processes (e.g., libraries, kernel data structures). It is a measure of the memory footprint of the process, which is useful for understanding how much memory a program requires to run and for identifying potential memory leaks. The maxrss value is typically reported by the getrusage() function, which is used in the provided code to measure memory usage.} ***

\section{Evaluation}
Performance studies were conducted on two crirteria: 
\begin{itemize}
    \item the number of generated hashes per second(Hashes/second)
    \item Memory usage 
\end{itemize}

The evaluation of three hashing algorithms, SHA256, SHA512, and BLAKE3 was conducted by varying the parallel threads from 1 to 8 using C programming language. The experiment was run for 500000 iterations while varying the input size on two machines: MacBook Pro and EC2 Large T2.2x Instance. The evaluation aimed to determine the effect of the number of parallel threads on the throughput of the hashing algorithms. The evaluation was conducted using the shared code snippets for each of the algorithms. The performance metrics were measured in terms of memory usage and throughput/ hashes per second. The results of the evaluation were analyzed to determine the optimal number of threads that would produce the highest throughput  for each of the algorithms. The results of the evaluation provide insights into the performance of the hashing algorithms under different parallel computing scenarios. 

The following graphs summarize the results of comparison of performance of these algorithms on Macbook pro(2020) . In the [fig. 6] it is shown the how many hashes are generated per second with 1 MB input and on single thread.  I used this a base benchmark and later varied the Input size \& the number of threads which can be noticed in former graphs.

\begin{figure}[htbp]
\centerline{\includegraphics{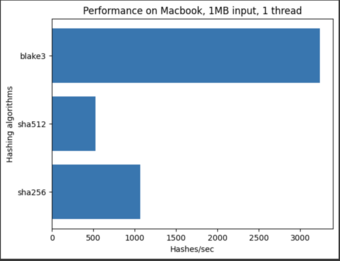}}
\caption{Performance on 1MB input size and 1 thread}
\label{fig}
\end{figure}

\begin{figure}[htbp]
\centerline{\includegraphics{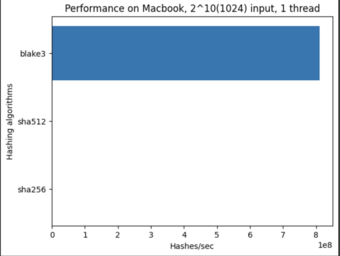}}
\caption{Performance on small input size \& 1 thread}
\label{fig}
\end{figure}
 \begin{figure}[htbp]
\centerline{\includegraphics{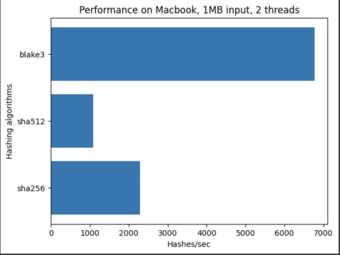}}
\caption{Performance on 1MB input size \& 2 thread}
\label{fig}
\end{figure}

\begin{figure}[htbp]
\centerline{\includegraphics{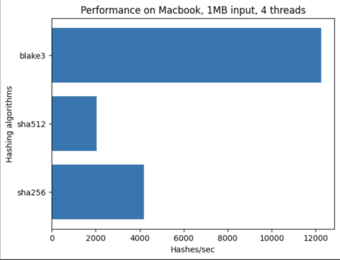}}
\caption{Performance on 1MB input size \& 4 threads}
\label{fig}
\end{figure}
\begin{figure}[htbp]
\centerline{\includegraphics{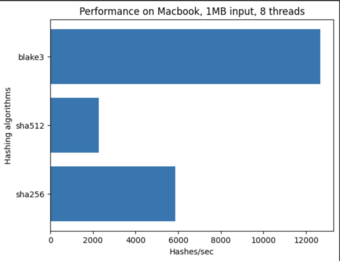}}
\caption{Performance on 1MB input size \& 8 threads}
\label{fig}
\end{figure}

my evaluation results show that Blake3 generally outperformed both SHA-512 and SHA-256 in terms of hashes generated per second on tested hardware platforms. Specifically, I found that Blake3 generated very high number of hashes compared to other two hashing algorithms.

\textbf{\textit{Memory Usage Evaluation - }}

In terms of memory usage, Blake3 is generally considered more efficient than SHA-256 and SHA-512. Blake3 is designed to use a smaller memory footprint than other popular hashing algorithms while still providing high performance and security.Blake3 uses a tree hashing algorithm that allows it to divide the input data into smaller block s and process them in parallel, which reduces the overall memory usage. In contrast, SHA-256 and SHA-512 use a sequential processing algorithm, which can require more memory to process large inputs.
However, it's worth noting that the memory usage of each hashing algorithm can depend on the specific implementation and the hardware platform. The amount of memory available on the system can also affect the performance and memory usage of each algorithm. Therefore, it's important to consider the specific use case and hardware platform when selecting a hashing algorithm based on its memory usage. 

In terms of memory usage, my results showed that Blake3 generally used less memory than SHA-512 and SHA-256, particularly when processing large input data sizes. if we closely look at the performance graphs, the difference of memory usage in general is very huge. 

\begin{figure}[htbp]
\centerline{\includegraphics{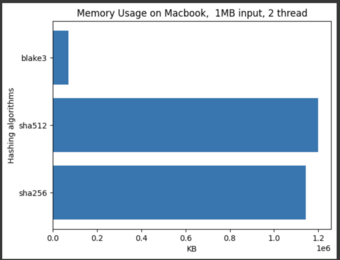}}
\caption{Memory Usage on 1MB input size \& 2 threads}
\label{fig}
\end{figure}

\begin{figure}[htbp]
\centerline{\includegraphics{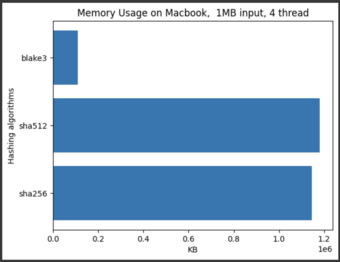}}
\caption{Memory Usage on 1MB input size \& 4 threads}
\label{fig}
\end{figure}

\begin{figure}[htbp]
\centerline{\includegraphics{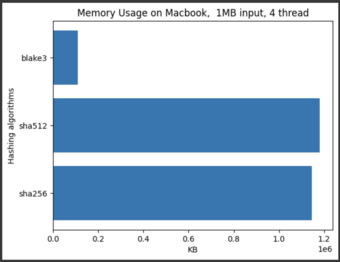}}
\caption{Memory Usage on 1MB input size \& 4 threads}
\label{fig}
\end{figure}
\begin{figure}[htbp]
\centerline{\includegraphics{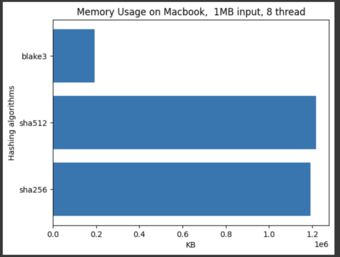}}
\caption{Memory Usage on 1MB input size \& 8 threads}
\label{fig}
\end{figure}

\textbf{\textit{Performance Evaluation on Linux EPYCBOX  - AMD EPYC 7501 32-Core Processor, 128 CPUs. }}
\begin{itemize}
    \item the number of generated hashes per second(Hashes/second)
    \item Memory usage 
    \item Caching performance 
    \item CPU utilisation
\end{itemize}

The evaluation aimed to compare the performance of three cryptographic hash algorithms, namely SHA-256, SHA-512, and Blake3, across several key metrics: CPU utilization, throughput, memory, and cache performance. The evaluation was conducted using a workload generated using \textbf{dd if=/dev/urandom of=synthetic\_data.txt bs=1G count=10 iflag=fullblock}
 consisting of 10 million iterations by varying the number of parallel threads to 1,2,4,16,32,64 \& 128. The input size was \textit{\textbf{32 bytes}} at a time.

I used\textit{\textbf{ perf}} monitoring tool to monitor the cpu, maximum resident size and also 
\textit{\textbf{/usr/bin/time -v ./program-name}} was helpful for the observation.

In terms of CPU utilization, Blake3 Outperfomed the other two, which can be seen in the figure as well. 

When it came to throughput, Blake3 showcased superior performance compared to SHA-256 and SHA-512. The inherent design optimizations of Blake3, such as SIMD instructions and parallelism, allowed it to efficiently process the workload and achieve higher throughput. Both SHA-256 and SHA-512, being older algorithms, exhibited slightly lower throughput, indicating that Blake3 outperformed them in terms of overall processing speed.

In terms of cache performance, Blake3 showcased notable improvements compared to SHA-256 and SHA-512. The algorithm's design and optimizations leveraged the CPU cache effectively, resulting in reduced cache misses and improved overall cache performance. In contrast, SHA-256 and SHA-512, lacking the same level of optimizations, may have experienced a slightly higher number of cache misses.

Overall, the evaluation suggests that Blake3 outperformed SHA-256 and SHA-512 in terms of throughput, CPU utilization, and cache performance.
\textbf{\textit{
CPU performance}} -  Blake3 is designed to take advantage of parallel processing capabilities available in modern CPUs. It utilizes SIMD (Single Instruction, Multiple Data) instructions to process multiple data elements simultaneously, which leads to efficient parallel execution and better CPU utilization. IT has efficient memory access. Blake3 can take advantage of hardware acceleration features available in some CPUs, like SIMD. 

\textbf{\textit{Caching performance - }} Blake3 is designed to have improved memory access patterns, which helps reduce cache misses. It utilizes a chunk-based processing approach, where the input data is divided into fixed-size chunks, allowing for better spatial locality. By accessing data elements in a contiguous and predictable manner, Blake3 can take advantage of cache lines more effectively. Blake3 makes efficient use of SIMD (Single Instruction, Multiple Data) instructions available in modern CPUs. SIMD instructions operate on multiple data elements simultaneously, enabling parallel processing and reducing memory access overhead. This can lead to improved cache utilization and reduced cache misses. 

\begin{figure}[htbp]
\centerline{\includegraphics{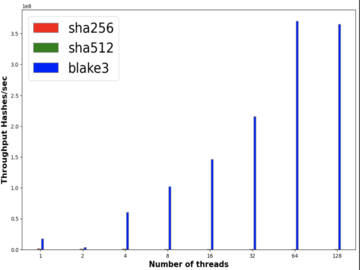}}
\caption{Throughput in terms of hashes/sec - 1e8 (y-axis) for threads 1,2,4,8,16,32,64,128}
\label{fig}
\end{figure}

\begin{table}[]
\begin{tabular}{lrlr}
\textit{Threads} & \multicolumn{1}{l}{\textbf{Blake3(hashes/sec)}} & \textbf{sha256(hashes/sec)} & \multicolumn{1}{l}{\textbf{sha512(hashes/sec)}} \\
1                & 17509761                                        & 1712624                     & 1138808                                         \\
2                & 3342002                                         & 1040261                     & 952239                                          \\
4                & 60316177                                        & 1094015                     & 1343171                                         \\
8                & 101868264                                       & 621361                      & 603409                                          \\
16               & 146389307                                       & 704976                      & 660650                                          \\
32               & 215461518                                       & 645208                      & 590431                                          \\
64               & 370288084                                       & 574635                      & 532922                                          \\
128              & 365483717                                       & 552507                      & 542918                                         
\end{tabular}
\caption{Table to show number of hashes generated per second in multiple experiments}
\end{table}
\begin{figure}[htbp]
\centerline{\includegraphics{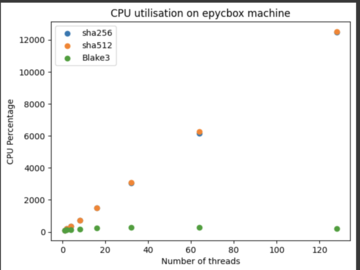}}
\caption{CPU utilization in terms of  percentage usage for threads 1,2,4,8,16,32,64,128}
\label{fig}
\end{figure}

\begin{figure}[htbp]
\centerline{\includegraphics{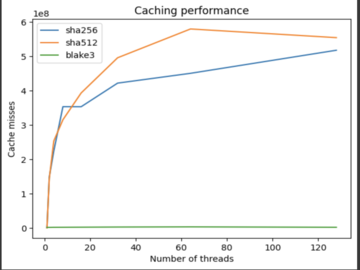}}
\caption{cache misses for evaluation for  threads 1,2,4,8,16,32,64,128}
\label{fig}
\end{figure}

\begin{figure}[htbp]
\centerline{\includegraphics{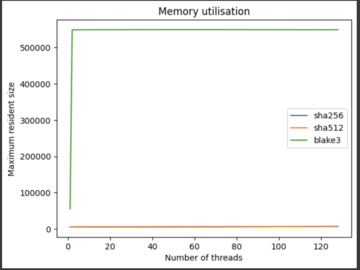}}
\caption{Memory in terms of Maximum resident size for threads 1,2,4,8,16,32,64,128}
\label{fig}
\end{figure}

\section{Related work}
A current field of research is the performance examination of hashing algorithms in the context of blockchain. Particularly, SHA-256 and SHA-512 have been used extensively in blockchain applications like Bitcoin and Ethereum and have undergone in-depth performance analysis. According to earlier research, SHA-256 and SHA-512 have quite high computing needs, especially when processing big volumes of data, and can use a lot of memory. Researchers have suggested different hashing algorithms to overcome these problems, such as Blake3, which is intended to be more memory-efficient and offer high-performance hashing for big data sets. With regard to blockchain, several studies have compared Blake3's performance to that of SHA-256 and SHA-512, with varying outcomes depending on the particular use.
I studied blake3 research paper, which motivated me to do the study on commodity machine, which helped me understand hashing algorithms better with respect to their implementation \& performance.  Things which were unique in my study was hardware, varied input size and different number of threads. My intention to do this study was to contribute an insights on commodity hardware like Mac machines, mostly researchers do performance analysis on finely configured machines, which indeed is useful but it's good to have studies on all the existing common machines to make pragmatic choice of the hardware from low level to high level configuration. 

\section{Conclusion}
Undertaking the project on performance evaluation of hashing algorithms on commodity hardware provided me with an extensive understanding of the popular hashing algorithms like SHA-256, SHA-512, and Blake3, including their performance characteristics. I also gained insights into the trade-offs between hashing speed, memory usage, and the algorithms' effectiveness under various input data sizes and hardware configurations. Becoming proficient with different hardware platforms like CPUs, GPUs, and benchmarking tools provided me with valuable experience in working with various hardware and improving my understanding of multithreading programming in C language. I also learned about different hardware issues and enjoyed troubleshooting them. Drawing meaningful insights from the performance data generated during the project improved my data analysis skills. Alongside learning about the various applications of hashing algorithms, I also delved into the specific use of hashing in blockchain, coupled with a practical application of multithreaded programming.

The project was a great learning outing throughout, it was a quite a success as I got to learn about numerous things but I still have experiments going on t2.2x large ec2 instance, and I faced more chanllenges on these machines as I ran few experiments I saw that sha512 outperforms sha256 in some cases, as a part of future work I am going to investigate why there  is performance difference, currently what I know is the performance of hashing algorithms can depend on various factors such as the CPU architecture, memory speed, cache size, and implementation details and depends on the block size for processing as well. So I am going to optimise the code, and get more concrete results for all the three algorithms. As per now, below table has the current findings on the t2.2x large instance, one of the table has base experiment as well. *** \textit{TBD for the SHA512, benchmark for this needs optimisation \& code changes, which work in progress I would pursue as my future work}

\begin{figure}[htbp]
\centerline{\includegraphics{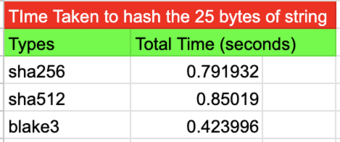}}
\caption{Table for the base experiment with a given string}
\label{fig}
\end{figure}

\begin{figure}[htbp]
\centerline{\includegraphics{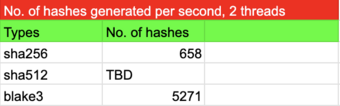}}
\caption{Figure shows number of hashes generated per second for 1 MB input \& 500000 iterations for hash generation}
\label{fig}
\end{figure}

\begin{figure}[htbp]
\centerline{\includegraphics{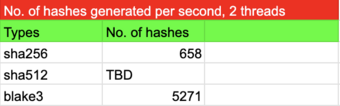}}
\caption{Figure shows number of hashes generated per second for 1 MB input \& 500000 iterations for hash generation}
\label{fig}
\end{figure}

\begin{figure}[htbp]
\centerline{\includegraphics{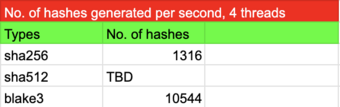}}
\caption{Figure shows number of hashes generated per second for 1 MB input \& 500000 iterations for hash generation}
\label{fig}
\end{figure}

\begin{figure}[htbp]
\centerline{\includegraphics{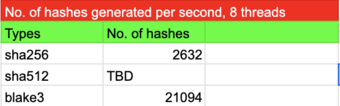}}
\caption{Figure shows number of hashes generated per second for 1 MB input \& 500000 iterations for hash generation}
\label{fig}
\end{figure}
\section*{Acknowledgment}
I would like thank Professor Ioan Raicu for his guidance \& valuable feedback throughout, It helped in making this project a success. I would also like to thank Alexandru Iulian Orhean for his valuable feedback for the presentation \& helping me in finding tools \& libraries for the development \&  execution of the benchmarks.

\vspace{12pt} 
\color{red}

\end{document}